\documentclass[aps,prb]{revtex4}
\usepackage{graphicx}
\usepackage{bm}
\usepackage{xfrac}
\usepackage{float}
\newcommand{\be}{\begin{equation}}
\newcommand{\ee}{\end{equation}}
\newcommand{\bea}{\begin{eqnarray}}
\newcommand{\eea}{\end{eqnarray}}
\newcommand{\beb}{\begin{eqnarray*}}
\newcommand{\eeb}{\end{eqnarray*}}

\begin{document}

\title{Spin and valley ordering of fractional quantum Hall states in
monolayer graphene}

\author{Ngoc Duc Le}
\author{Thierry Jolicoeur}
\affiliation{Universit\'e Paris-Saclay, CNRS, CEA, Institut de Physique Th\'eorique, France}

\date{November 30th, 2020}
\begin{abstract}
We study spin and valley ordering in the quantum Hall fractions in monolayer graphene
at Landau level filling factors $\nu_G=-2+n/3$ $(n=2,4,5)$.
 We use exact diagonalizations on the spherical as well as toroidal
 geometry by taking into account the effect of realistic anisotropies that break the spin/valley
 symmetry of the pure Coulomb interaction. 
We also use a variational method based on eigenstates of the fully $SU(4)$ 
symmetric limit.
For all the fractions we study there are two-component states for which
the competing phases are generalizations of those occurring at neutrality $\nu_G=0$.
They are ferromagnetic, antiferromagnetic, charge-density wave and K\'ekul\'e phases,
depending on the values of Ising or XY anisotropies in valley space.
The varying spin-valley content of the states leads to ground state quantum numbers
that are different from the $\nu_G=0$ case.
For filling factor $\nu_G=-2+5/3$ 
there is a  parent state in the $SU(4)$ limit which has a flavor content 
$(1,1/3,1/3,0)$ where the 
two components that are one-third filled form a two-component singlet. The addition 
of anisotropies
leads to the formation of new states that have no counterpart at $\nu_G=0$. While some 
of them
are predicted by the variational approach, we find notably that negative Ising-like 
valley anisotropy
leads to the formation of a state which is a singlet in both spin and valley
space and lies beyond the reach of the variational method. 
Also fully spin polarized two-component states at $\nu=-2+4/3$ and $\nu=-2+5/3$
display an emergent $SU(2)$ valley symmetry because they do not feel
point-contact anisotropies.
We discuss
implications for current experiments concerning possible spin transitions.
\end{abstract}
\maketitle
\section{introduction}
The fractional quantum Hall effect~\cite{Tsui} (FQHE) occurs in 
two-dimensional electron
systems in a strong perpendicular field and is characterized notably by a gap for all
charged excitations at some fractional filling $\nu$ of the Landau levels.
Such gaps are due to electron-electron Coulomb interactions. In this regime
there are excitations with fractional charges and also statistics, Abelian or even 
non-Abelian in some cases. There are possible opportunities for fabrication of new
electronic devices
using the non-Abelian statistics for quantum computation~\cite{Nayak} or the 
interlayer phase coherence
that occurs in bilayer systems for dissipationless devices~\cite{Eisenstein}. 

For many years after its discovery the FQHE has been studied almost only in the special
two-dimensional electron gases formed in GaAs/GaAlAs heterojunctions.
In this set-up the electron Land\'e $g$-factor and the dielectric constant 
of the host material conspire to reduce the Zeeman energy with respect to the Coulomb
energy scale. As a consequence the spin degree of freedom in the lowest Landau level
cannot be considered as frozen by the external magnetic field and FQHE ground states
as well as their excited states are not always fully spin-polarized. This leads notably
to quasiparticles that have nontrivial spin textures called
skyrmions~\cite{Sondhi,Moon}.
The study of electron gases in materials such as AlAs~\cite{shayegan} added an extra degeneracy
due to the relevance of several valleys for the electronic states.

The discovery of monolayer graphene has opened an even richer 
arena~\cite{Du,Bolotin,Ghahari,Dean-MLG1,Young-MLG2,Feldman-MLG3,Feldman-MLG4,Hunt,Young-MLG4,Amet-MLG5,Zibrov-MLG6,Polshyn-MLG7,Chen-MLG8,Zeng-MLG9,Zhou-MLG10,Yang-MLG11,Zhou-MLG12}
for the FQHE
since there is also an additional two-fold valley degeneracy that comes to play.
The central $N=0$ Landau level of monolayer graphene is  approximately fourfold
degenerate because of spin and valley degrees of freedom
and is partially filled for a range of filling factor $-2\leq\nu_G\leq+2$.
Integer quantum Hall states~\cite{dassarmayang} also appear at fillings 
$\nu_G=0,\pm 1$ and this is an instance
of quantum Hall ferromagnetism~: Coulomb interactions lead to gap opening also
at these integer filling factors.

At neutrality $\nu_G=0$ theory predicts~\cite{JJung,Kharitonov} competing phases 
with various patterns of
spin and valley ordering. One can have ferromagnetic, antiferromagnetic, K\'ekul\'e
or charge-density wave states. While these states are degenerate if one makes
the approximation of full $SU(4)$ symmetry in spin/valley space, anisotropies will
select one of these states in real samples. Changing parameters like the ratio 
of Zeeman
energy to Coulomb energy can induce transitions between such ground states.
An appealing scenario has been presented to explain a transition from an insulator
to a quantum spin Hall state as the transition between the canted antiferromagnetic 
state and the ferromagnetic state.
When the central Landau level is partially filled we expect formation of FQHE states
and they will have also some pattern of spin and valley ordering.
The fully $SU(4)$ symmetric case has been explored in many works~\cite{Apalkov,Toke2006,Toke2007,Shibata1,Shibata2,Toke2,Papic,Balram2015,Balram2015b,Balram2015c}.
Inclusion of anisotropies has been studied by Abanin et al.~\cite{Abanin} 
at the level of wavefunctions
and an extension of Hartree-Fock theory has been also proposed by Sodemann and 
MacDonald~\cite{Inti}. While the results of mean-field theory have been confirmed 
by exact
diagonalizations~\cite{Wu2014} for $\nu_G=0$ this has not been checked for the 
fractional quantum Hall states.

In this paper we introduce a set of variational wavefunctions constructed out of
exact eigenstates of the $SU(4)$ symmetric limit. The wavefunctions are dressed by
a spin and flavor configuration who is determined by minimizing the energy.
The anisotropy energy can be expressed solely in terms of the pair correlation function 
of the parent symmetric state. This approach generalizes the mean-field treatment of
quantum Hall ferromagnetism. Within this framework we obtain the phase diagrams
for the fractions $\nu_G=-2+n/3$ ($n=2,4,5$). We also perform extensive exact
diagonalizations in the spherical~\cite{Haldane,Fano} and torus geometry to check the
validity of the
variational approach. 
For fractions $\nu_G=-2+n/3$ ($n=2,4,5$) there are two competing ground states
in the $SU(4)$ limit. At $\nu_G=-2+2/3$ we have the fully polarized particle-hole
partner of the $\nu_G=-2+1/3$ as well as a singlet state with two-component
structure, well known from previous studies~\cite{Ashoori}.
At $\nu_G=-2+4/3$ the competing states are the two-component particle-hole
partner of the singlet with spin-valley content $(2/3,2/3,0,0)$
and a state with a filled shell by addition to the $\nu=1/3$ Coulomb ground state
with spin-valley content $(1,1/3,0,0)$.
At $\nu_G=-2+5/3$ we have a two-component state $(1,2/3,0,0)$
where the $2/3$ state is the fully polarized state
and there is also a three-component state $(1,1/3,1/3)$ that is
built from a filled Landau level plus a spin-singlet $2/3$ state
as pointed out by Sodemann and MacDonald~\cite{Inti}
The phase diagrams from the variational method are in agreement
with the diagonalization results except for the fraction $\nu_G=-2+5/3$
where a large portion of the phase space is occupied by a correlated singlet state
that cannot be reproduced by our set of wavefunctions. To clarify its structure
we compute the pair correlation function for all combinations of spin/valley indices
and find that there is an enhanced probability for pair formation of opposite spin.
We discuss possible ways of favoring this new phase in real samples of graphene.
While the regions of stability of the different patterns in spin and valley 
configurations are correctly predicted variationally the quantum numbers
are different in several important cases involving antiferromagnetism.
Also we observe that fully polarized phases for the states $(1,1/3,0,0)$ and
$(1,2/3,0,0)$ still form $SU(2)$ valley multiplets even though the Hamiltonian 
does not have such a symmetry. This is due to the special point-contact form
of the anisotropies.
For all the fractions we have studied there are in general spin transitions
when one varies the magnetic field. Multicomponent states are preferred
at small Zeeman energies.

In section II we describe the Landau levels of monolayer graphene. In section III
we introduce the Coulomb interactions and anisotropies that govern the spin/valley
ordering of quantum Hall states. Section IV contains  a discussion of the $SU(4)$
representations that we use in our exact diagonalization studies as well as the
formulation of trial wavefunctions for fractional states with various spin/valley
configurations. Section V discuss the integer quantum Hall states at $\nu_G=0,\pm 1$.
Section VI give results for fractions $\nu_G <-1$. In section VII we give evidence
for similar physics at $\nu_G=-2+4/3$ and $\nu_G=0$. The phases for the 
three-component state 
at $\nu_G=-2+5/3$ are discussed in section VIII.  We give a simplified treatment 
of spin transitions in section  IX and section X contains our conclusions. 

\section{graphene Landau levels}
Monolayer graphene has a simple hexagonal lattice structure that leads to a 
massless Dirac fermion
spectrum close to the neutrality point. When applying a perpendicular magnetic field
there is formation of relativistic Landau levels with energies~:
\be
E_N=sign(N)\frac{\hbar v_F }{\ell}\sqrt{2|N|} +\frac{1}{2}g\mu_B B \sigma,\quad N=0,\pm1, \pm 2, \dots
\label{GLL}
\ee
where $\ell=\sqrt{\hbar/eB}$ is the magnetic length, $v_F$ is the velocity of 
the relativistic Dirac fermions, $g$ is the Land\'e factor which is equal to $2$
because spin-orbit coupling is negligible in graphene, $\mu_B$ is the Bohr magneton
and $\sigma=\pm 1$ the spin projection onto the magnetic field direction.
All these Landau levels are twice degenerate due to the presence of two valleys $K$ and $K^\prime$ in
the graphene Brillouin zone.
In this paper we concentrate on the physics that takes place in the zero-energy $N=0$ central
Landau level for fractional filling factors.
The $N=0$ Landau level wavefunctions have the same spatial dependence as the
non-relativistic two-dimensional electrons. For the $N=0$ manifold of Landau states the
valley index
translates exactly into
a sublattice index i.e. a valley $K$ Landau state has nonzero amplitude only on one
sublattice
 $A$ while the other valley $K^\prime$ is entirely concentrated on the other 
 sublattice $B$.
An important consequence is that a sublattice potential difference $\Delta_{AB}$ is
equivalent
to a  Zeeman field acting on the valley degree of freedom i.e. lifting the 
degeneracy between
the valley
degrees of freedom. Such a potential is not expected for suspended graphene samples 
but is
induced notably by a hexagonal boron nitride (hBN)
substrate when it is geometrically aligned with the graphene layer.

The energy levels from Eq.(\ref{GLL}) lead to integer quantum Hall effects at
graphene filling factors $\nu_G=\pm 2, \pm 6, \pm 10,\dots$. However even in 
the integer
quantum Hall regime it was discovered that there are also states for $\nu_G=0,\pm 1$
within the $N=0$ Landau level. These are instances of the general phenomenon of 
quantum Hall ferromagnetism~\cite{dassarmayang}. Indeed the general arguments 
for appearance 
of quantum Hall ferromagnetism carry out in the case of graphene. We discuss the 
case of fillings $\nu_G=0,\pm1$ below.
It is convenient to redefine the filling factor by writing $\nu=2+\nu_G$
so that the $N=0$ LL now spans the range $0\leq\nu\leq+4$. 
The global particle-hole symmetry maps the filling factor $4-\nu$ onto $\nu$ so that 
it is enough to restrict
our study to the range $0\leq\nu\leq+2$.

\section{interactions and anisotropies in monolayer graphene}
 We now discuss the effective electron-electron interactions within
 the $N=0$ Landau level of monolayer graphene. The Coulomb interaction
 is $SU(4)$ symmetric to an excellent approximation. It leads to an
 energy scale which is constructed out of the magnetic length
 $E_C=e^2/(\epsilon\ell)$ where $\ell=\sqrt{\hbar/(eB)}$ and $\epsilon$
 the dielectric permittivity of the system. The unitary $SU(4)$ symmetry mixing
 spin and valley degrees of freedom is however reduced by several phenomena.
 It has been proposed to encapsulate these splittings into the following
 Hamiltonian~\cite{Alicea,Kharitonov} acting only onto the valley degrees of freedom~:
\be
\mathcal{H}_{aniso}=\sum_{i<j}
\left[g_\perp (\tau^x_i \tau^x_j+\tau^y_i \tau^y_j) + g_z \tau^z_i \tau^z_j \right]
\delta^{(2)}(\mathbf{r}_i-\mathbf{r}_j),
\label{aniso}
\ee
where the $\tau^\alpha_i$ Pauli matrices operate only in valley space. 
We note that this form of the interaction is local in real space so it is not felt by 
quantum states
that vanish when two particles coincide in space.
This is the case notably for the $\nu=1/3$ Laughlin wavefunction.
This simple model reminiscent of the $XXZ$ model for magnetic systems
has several desirable features, notably it describes the metal-insulator transition
observed at $\nu=0$ when tilting the magnetic field away from the direction perpendicular
to the sample as we discuss below~\cite{Young-MLG4}.
It is convenient to parameterize the two coefficients $g_{\perp,z}$ with an 
angular variable $\theta$~:
\be
g_\perp = g\cos\theta , \quad g_z=g\sin\theta ,
\ee
in addition to an overall strength $g$ which we take as positive.
It is convenient to convert the parameters $g_{\perp,z}$ into two separate 
energy scales~:
\be
u_\perp = \frac{g_\perp}{2\pi \ell_B^2}, \quad u_z = \frac{g_z}{2\pi \ell_B^2}.
\ee
One can define dimensionless strengths of anisotropies by factoring out the Coulomb 
energy scale~:
\be
\tilde g = (g/\ell_B^2)/(e^2/(\epsilon\ell_B))
\ee

One-body energy level splitting occurs through the Zeeman effect that acts
on the spins and sublattice splitting onto the valley indices~:
\be
\mathcal{H}_{1body}=-\epsilon_Z \sum_i S^z_i + \Delta_{AB}\sum_i T^z_i .
\ee
The Zeeman energy $\epsilon_Z$ is $g\mu_B B_{T}$ where $B_T$ is the total field and the 
Land\'e factor $g=2$ since spin-orbit coupling is negligible in graphene. We note that the 
direction of the field is arbitrary due to spin
rotation invariance. On the contrary the sublattice symmetry breaking takes
place between the two valleys. In the case of the commonly used hBN substrate
typical values of the splitting $\Delta_{AB}$ are of the order of 10 meV
and are magnetic field independent.
 
The special Hamiltonian Eq.(\ref{aniso}) has some symmetry properties that are independent 
of the filling factor.
There are symmetries like those of the XXZ Heisenberg Hamiltonian well known in the field
of quantum magnetism. Notably one can always perform rotations in valley space around 
the $z$ axis.
These form a $U(1)$ symmetry group leading to the conservation of $T_z$, the projection of 
the isospin onto the $z$ axis.
When $u_z=u_\perp$ ($\theta=\pi/4,\pi+\pi/4$) there is invariance under full rotation in 
valley space with a group
$SU(2)$. Beyond these obvious symmetries, we note that for $u_\perp=0$ 
($\theta=\pi/2,3\pi/2$) one can make spin rotations
independently in each valley so there is a symmetry $SU(2)_K\times SU(2)_{K^\prime}$.
There is also an additional $SO(5)$ symmetry  when $u_z+u_\perp=0$ 
($\theta=3\pi/4,7\pi/4$) which is not obvious in 
the present formulation~\cite{Wu2015}.

It is important to keep in mind that the anisotropic Hamiltonian Eq.(\ref{aniso}) 
has no
fundamental significance. It should be viewed as the most relevant perturbation beyond
the $SU(4)$ symmetric Coulomb interaction. In the real world there will be weaker
anisotropic interactions between electrons with non-zero relative angular momentum.
We also note that the magnitude of the couplings $g_z,g_\perp$ is still uncertain.
Constraints on their values come from the metal-insulator transition in tilted field
observed at neutrality~\cite{Young-MLG2}.

\section{integer fillings $\nu_G =0,\pm 1$}
If we consider the integer quantum Hall state at $\nu=1$ then there is a set of 
exact eigenstates
of the Coulomb problem with a closed form given by a single Slater determinant~:
\be
|\Psi_{\nu=1}\rangle=\prod_m c^\dag_{m\alpha} |0\rangle,
\label{nu1wave}
\ee
where $|\alpha\rangle$ is an arbitrary four component vector in spin-valley space.
The integer $m$ is the index of the orbital Landau level and the product runs over
all values of $m$, corresponding to complete filling of the level.
This state is an exact eigenstate provided one neglects Landau level mixing. This 
fact is 
aptly called quantum Hall ferromagnetism~\cite{dassarmayang}.
The arbitrariness in the the direction of $|\alpha\rangle$ is fixed presumably by 
lattice effects
beyond the simple continuum models we consider~\cite{Alicea}. Notably such 
a wavefunction 
Eq.(\ref{nu1wave})
vanishes when electrons coincide in real space and so it is insensitive to 
perturbations like the anisotropies of Eq.(\ref{aniso}).

\begin{figure}[H]
\centering
 \includegraphics[width=0.3\columnwidth]{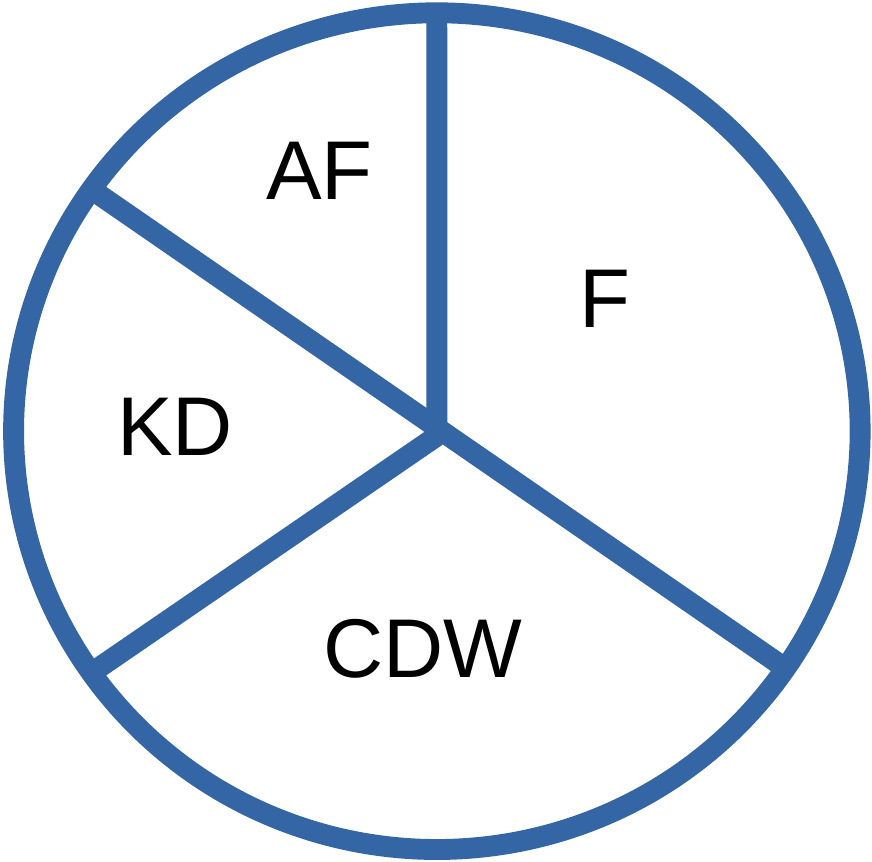}
 \caption{Phase diagram for neutrality $\nu=2$. The various ground states are displayed as a function
 of the anisotropy angle $\theta$ that varies $g_z=g\sin\theta$ and $g_\perp=g\cos\theta$.
 For small enough anisotropy energy, the precise value of the overall
 energy scale $g$ is irrelevant to the  phase competition.
 There are four phases and four first-order phase transitions between them, occurring at high-symmetry points.}
 \label{nu0phase}
\end{figure}

We now discuss the special case of graphene neutrality $\nu=2$. At this filling factor
there are exactly two filled 
Landau levels. In the fully symmetric $SU(4)$ limit one notes that a Slater 
determinant is 
an exact eigenstate~:
\be
|\Psi_{\nu=0}\rangle=\prod_m c^\dag_{m\alpha} c^\dag_{m\beta}|0\rangle,
\label{nu0wave}
\ee
where the orbital index $m$ (whose precise definition is geometry dependent) takes all allowed values
corresponding to complete integer filling, and $\alpha,\beta$ are two orthogonal vectors
in four-dimensional spin-valley space spanned by 
$\{|K\uparrow\rangle,|K\downarrow\rangle,
|K^\prime\uparrow\rangle,|K^\prime\downarrow\rangle\}$. 
Due to the $SU(4)$ symmetry these two vectors are arbitrary.
The anisotropies will induce energy differences among all possibilities and determine 
the ground state
spin-valley ordering. The mean-field treatment is performed by taking the 
expectation value 
of the
anisotropy Hamiltonian Eq.(\ref{aniso}) in the Slater determinant of 
Eq.(\ref{nu0wave}).
The result is given by a functional of the two vectors $\alpha$ and $\beta$~:
\be
\epsilon_a=\frac{1}{N_{\phi}} 
\langle\Psi_{\nu=0}| \mathcal{H}_{aniso}|\Psi_{\nu=0}\rangle
=\sum_{i=x,y,z} u_i [ tr(P_\alpha \tau_i)tr(P_\beta \tau_i)
-tr(P_\alpha \tau_i P_\beta \tau_i)]
\label{trialenergy}
\ee
where we have defined 
the anisotropy energy per flux quantum $\epsilon_a$, the
projectors onto the trial vectors~:
\be
P_\alpha=|\alpha\rangle\langle\alpha|,P_\beta=|\beta\rangle\langle\beta|,
\ee
and the couplings are $u_x=u_y=u_\perp$, $u_z$.
By minimization of this functional
one finds four phases in the absence of Zeeman energy that are displayed in 
Fig.(\ref{nu0phase}).
There is a ferromagnetic phase (F) with $\alpha=|K\uparrow\rangle,\beta=|K^\prime\uparrow\rangle$
which is stabilized in the range $-\pi/4<\theta<+\pi/2$. One finds also
an antiferromagnetic phase (AF) with $\alpha=|K\uparrow\rangle,\beta=|K^\prime\downarrow\rangle$
preferred in the range $\pi/2 < \theta  < 3\pi/4$. Note that this state is an
antiferromagnet both in spin space and valley space.
The K\'ekul\'e state (KD) has $\alpha=|\mathbf{n}\uparrow\rangle,\beta=|\mathbf{n}\downarrow\rangle$
where $\mathbf{n}$ is a vector lying in the XY plane of the Bloch sphere for valley degrees of freedom~:
$\mathbf{n}=(|K\rangle+\mathrm{e}^{i\phi}|K^\prime\rangle)/\sqrt{2}$ where $\phi$ is an arbitrary angle in the valley XY plane.
This state is a spin singlet but a XY valley ferromagnet.
It is preferred in the range $3\pi/4<\theta<5\pi/4$. Finally
 the charge density wave state (CDW) defined by 
 $\alpha=|K\uparrow\rangle,\beta=|K\downarrow\rangle$.
This state is a spin singlet but a valley ferromagnet. Since the valley index 
coincides with the sublattice index
in the central Landau level, this state has all the charge  density on one 
sublattice and would be favored by a substrate
breaking explicitly the sublattice symmetry like hexagonal boron nitride. It 
requires the range
$5\pi/4<\theta<7\pi/4$ as it is stabilized by negative valley interactions along 
$z$ direction.

All transitions between these various states are first-order within mean-field theory
and this is confirmed by exact diagonalizations~\cite{Wu2014}. Notably all transitions
take place on the points with extra symmetries beyond the $U(1)$ valley conservation.
The spin quantum number of the ground state for a finite system as studied by exact
diagonalization is given by $S=0$ i.e. spin-singlet states for AF, KD and CDW phases
while the F phase has maximal spin $S=N_e/2$. Concerning valley quantum numbers
there are three phases with $T^z=0$~: F, AF, and KD while CDW has maximal valley
polarization $T^z=N_e/2$.

With nonzero Zeeman energy the most notable change is that the AF phase becomes 
canted (CAF).
The spins prefer to have antiparallel orientation in the plane perpendicular to the
field direction and a small projection onto this direction. Increasing the Zeeman energy
leads to greater ferromagnetic character and ultimately a phase transition from CAF to 
the F
phase. The F phase has the very special characteristic of having conducting edge modes
contrary to the CAF phase~\cite{ALL}. 
The tilted-field experiments of ref.(\onlinecite{Young-MLG2}) enable the variation
of the ratio of Zeeman vs. Coulomb energy scales. Since these is a metal-insulator
transition in this set-up a natural explanation is that they observe the CAF-F
transition.
To be consistent with this scenario one deduces bounds on anisotropies~:
\be
u_\perp\simeq -10\epsilon_Z, \quad u_z+u_\perp > 0.
\label{anisobounds}
\ee

\section{SU(4) representations and model wavefunctions}
\label{irreps}

In this section we describe how we use the theory of irreducible representations 
of the $SU(4)$
group in our exact diagonalization studies.
Energy levels of a $SU(4)$-invariant Hamiltonian are in general degenerate 
and form irreducible
representations (irreps) of the symmetry group. These irreps are
in one-to-one correspondence with Young tableaux that consist of 
three rows of boxes with $L_1$ boxes on the first row, $L_2$ boxes on the 
second row and $L_3$
on the third row with $L_1\geq L_2\geq L_3$. The dimension of such an irrep is given by~:
\be
\mathcal{D}(p_1,p_2,p_3)=\frac{1}{12}(p_1+1)(p_2+1)(p_3+1)
(p_1+p_2+2)(p_2+p_3+2)(p_1+p_2+p_3+3),
\ee
where we have used the positive integers $p_1=L_1-L_2$, $p_2=L_2-L_3$,
$p_3=L_3$. It is also convenient to view such an irrep as a collection
of irreps of $SU(2)$ subgroups that operate only on subsets of the basis states. 
We will use
repeatedly the pure spin subgroup generated by
$S^\alpha=\sigma^\alpha \otimes 1$ and the pure isospin subgroup generated
by $T^\alpha= 1\otimes \tau^\alpha$ acting on the valley degrees of freedom. They can be
conveniently complemented by a third $SU(2)$ subgroup
with generators $N^\alpha=\sigma^\alpha\otimes \tau^z$ that
is inspired by N\'eel antiferromagnetic order.
Although there are six independent such $SU(2)$ subgroups only these three are convenient
because we will consider the additional symmetry-breaking perturbation given by
Eq(\ref{aniso}). A state belonging to a given
$SU(4)$ irrep
can be taken as an eigenstate of $(S^z,T^z,N^z)$ simultaneously
because these generators commute. Contrary to the simpler case
of the $SU(2)$ group there is in general more than one state characterized
by the three values $(S^z,T^z,N^z)$.

The $SU(4)$ symmetry implies that the second-quantized Hamiltonian of the Coulomb interaction
preserves the number of particles of each species $n_i$ with definite spin and 
valley so one may
impose separately these numbers $(n_1,n_2,n_3,n_4)$
and perform the diagonalization in the sector defined by these fixed numbers.
The identification of an irrep is made by finding a highest weight state. 
This concept is
familiar in the $SU(2)$ case.
Indeed if one finds an energy value by imposing a given $S^z=M$
one does not know the total spin by this sole observation~:
we only know that the total spin of this level is larger than
$M$. One has to find eigenenergies in sectors with $S^z=M+1$
then $S^z=M+2$ and so on up the point of disappearance of the energy level.
The value of highest $S^z$ containing a given energy is then equal to its
\textit{total} spin. A generalized reasoning applies in the $SU(4)$ case.
If an energy eigenstate is found in some $(n_1,n_2,n_3,n_4)$
sector one has to track it among states obtained by acting with raising operators 
till one finds a highest-weight state
annihilated by all such operations. This is more computationally demanding than in 
the $SU(2)$ case. For a value of the highest weight $(S^z,T^z,N^z)=(m_1,m_2,m_3)$ 
the $SU(4)$ irrep is characterized by
integers $p_1=m_2-m_3, p_2=m_1-m_2,p_3=m_2+m_3$ and the corresponding values of the 
lengths of Young tableau rows $L_1,L_2,L_3$.

The picture that underlies our investigations is that eigenstates are essentially
$SU(4)$ multiplets whose
degeneracy
is lifted only perturbatively by the anisotropies in Eq.(\ref{aniso}). 
This is sensible as 
long as the anisotropy energies are not
too large with respect to the Coulomb energy scale. Present experiments seem 
to be compatible
with this point of view. It is important to note that the anisotropy energies are likely
larger than the Zeeman energy in current experiments~\ref{anisobounds}.
All these symmetry considerations above are valid independently of the geometry we use.
We make use of the spherical geometry~\cite{Haldane} as well as of the torus geometry~\cite{HaldaneTorus}. The ground state of a FQHE system is expected
to be uniform in space so on a sphere it should have zero total angular momentum
and on the torus geometry where one can define magnetic many-body translations
it should have also zero many-body momentum $K_x=K_y=0$.
While the pure Coulomb problem has the complete $SU(4)$ symmetry, the anisotropic case
admit only spin $T^z$ conservation. For practical reasons we implement only $S^z$
and $T^z$ conservation.

We now discuss trial wavefunctions that can be used to describe the fractional 
quantum Hall
states with spin and valley degrees of freedom. We first make several remarks that 
apply to
the $SU(4)$ symmetric limit. Note that
any $SU(2)$ Coulomb eigenstate is also an $SU(4)$ eigenstate. Only the degeneracy will
change. If we have some $SU(2)$ Coulomb eigenstate constructed with two 
spin-valley vectors $\alpha$ and 
$\beta$ and filling factor $\nu_{\alpha\beta}$ then we can glue a filled $\nu=1$ complete
Landau level for any vector $\gamma$ and obtain another exact eigenstate of 
the Coulomb interaction
with now filling factor $\nu=1+\nu_{\alpha\beta}$~:
\be
\Psi= \{\prod_m c^\dag_{m\gamma}\} \hat{\Psi}^\dag_{\alpha\beta} |0\rangle ,
\label{product}
\ee
where the operator $\hat{\Psi}^\dag_{\alpha\beta}$ creates the $SU(2)$ state with 
two components.
The energy of this new state is now given by~:
\be
E_{1+\nu_{\alpha\beta}}=E_1 + E_{\nu_{\alpha\beta}}, \quad E_1=-\sqrt{\pi/8}E_C,
\ee
where $E_1$ is the energy of the completely filled level.
In first-quantized language this gluing operation is the multiplication
by the Vandermonde determinant of the $\nu=1$ state.
Note that this gluing operation also works if we use a single component state
$\hat{\Psi}_{\alpha}$ instead of $\hat{\Psi}_{\alpha\beta}$.

In addition to the full particle-hole symmetry mapping $\nu$ to $4-\nu$
one can also perform a particle-hole symmetry on only two flavours
mapping $(\nu_1,\nu_2)$ to $(1-\nu_1,1-\nu_2)$ and the total filling
is then transformed from $\nu$ to $2-\nu$. Under this operation
the energy becomes~:
\be
E_{2-\nu}=E_{\nu}+2(1-\nu)E_1 .
\label{pph}
\ee
These mappings Eqs(\ref{product},\ref{pph}) allow us to identify at least some of the
eigenstates found by exact diagonalization. Of course there are also 
multicomponent states
that cannot be generated from the 2-component case. Some of them have been found
at filling factor $\nu=2/3$ in a previous study~\cite{Wu2015}.
It is possible also to use these mappings to construct trial wavefunctions once we have
a one or two-component state as obtained for example by the composite 
fermion construction~\cite{Jain,Jainbook}. For the fractions we study, which are 
most prominent
in experiments we cannot use known multicomponent generalizations~\cite{Halperin,Allan}
of the Laughlin wavefunction because they lead to more complicated fractional fillings.

We adhere to the view that anisotropies are not strong enough to destroy the 
Coulomb correlations
in a given trial state. It means that the small symmetry-breaking perturbations 
Eq.(\ref{aniso}) will choose the orientation of the free vectors $\alpha,\beta,\gamma$
in a trial state like those of Eq.(\ref{product}). Sodemann and MacDonald~\cite{Inti} have
proposed an approximate scheme based on an extension of Hartree-Fock theory to estimate
the anisotropy energy as a function of the free vectors in a trial state like 
Eq.(\ref{product}). 
We note that it is fact feasible to compute directly the expectation value 
of the Hamiltonian for anisotropies in the trial states, bypassing any Hartree-Fock like approximation.
Since the anisotropic interactions are purely point-like, the expectation value
of the Hamiltonian Eq.(\ref{aniso}) can be expressed in terms of the pair correlation
function at the origin $g_{\alpha\beta}(0)$ generalizing the formula for $\nu=2$~:
\be
\epsilon_a=\sum_{i=x,y,z} u_i \sum_{\alpha\beta}\, g_{\alpha\beta}(0)\, [ tr(P_\alpha \tau_i)tr(P_\beta \tau_i)
-tr(P_\alpha \tau_i P_\beta \tau_i)],
\label{anisoenergypair}
\ee
where the pair correlation function is that of the trial wavefunction.
We define a more compact notation~:
\be
\epsilon_a=\sum_{\alpha\beta}\, g_{\alpha\beta}(0)\, 
\mathcal{F}_{\alpha\beta},
\label{anisopair}
\ee
where now the sum over $\alpha,\beta$ runs over all values involved in
the trial wavefunction. The sum runs only over distinct values due to the Pauli principle 
($g_{\alpha\alpha}(0)=0$). If we consider trial states obtained by gluing a filled
shell like in Eq.(\ref{product}) then there are no non-trivial correlations between the
completely filled shell and the other electrons~: $g_{1\alpha}(0)=\nu_\alpha$.
The case with two filled Landau levels $\nu_\alpha=\nu_\beta=1$ gives the previous
formula for the anisotropy energy Eq.(\ref{trialenergy}). 
If we take a single-component state
with filling $\nu_2$ and glue a $\nu=1$ shell then we obtain an energy which is
simply the $\nu=2$ formula multiplied by a $\nu_2$ factor. Hence
without any further calculation we can be sure that the phase diagram is
identical to that of the $\nu=2$ and is given in Fig.(\ref{nu0phase})
with the same set of spin-valley vectors described above for the $\nu=2$ case.
However since the number of electrons is different in the two components,
the quantum numbers of the ground state are different from those of the $\nu=2$ case.

We now discuss the case with three occupied flavors with content $(1,\nu,\nu)$.
The trial state is now a two-component state with flavor content $(\nu,\nu)$
with a filled shell $\nu=1$ glued onto it.
The anisotropy energy is now given by~:
\be
\epsilon_a^{(1,\nu,\nu)}=\nu [\mathcal{F}_{\alpha\beta}+\mathcal{F}_{\alpha\gamma}]
+g_{\beta\gamma}(0)\mathcal{F}_{\beta\gamma}
\ee
For the trial states we consider, the partially filled states will involve
the two-component $\nu=2/3$ and $\nu=2/5$ spin-singlet states that have  a very small
pair correlation between different spin values $g_{\alpha\beta}(0)\approx 10^{-3}$.
This very small number only slightly change the phase boundaries and can be
safely discarded.

For matrix elements between two states $|\mathbf{t}_1\mathbf{s}_1 \rangle$,
$|\mathbf{t}_2\mathbf{s}_2 \rangle$ (so without spin-valley entanglement) the 
matrix element $\mathcal{F}_{12}$ is given by~:
\be
\mathcal{F}_{12} =\frac{1}{2}(1-\mathbf{s}_1\cdot\mathbf{s}_2)[\sum_i u_i t_{1i}t_{2i}]
-\frac{1}{2}(1+\mathbf{s}_1\cdot\mathbf{s}_2)
\frac{1}{2}(1-\mathbf{t}_1\cdot\mathbf{t}_2)[\sum_i u_i ],
\ee
which allow us to obtain the variational energies for all cases of concern.

When discussing trial states we omit the unoccupied spin-valley states when displaying
the component structure of the wavefunction i.e. $(1,1/3,0,0)$ is written
simply $(1,1/3)$. However when describing an irrep we use all four components
i.e. $(7,4,0,0)$ stands for an $SU(4)$ irrep defined by its highest weight.
\section{Fractions for $\nu <1$}
\subsection{The $\nu=1/3$ state}
\label{onethird}

At filling factor $\nu=1/3$ the one-component Coulomb ground state
is an exact eigenstate of the $SU(4)$ symmetric case and it is a member
of the irrep with highest weight $(N_e,0,0,0)$. This means that the spatial part 
of the wave function is
multiplied by a fully symmetrized wavefunction with all electrons in the same
spin/valley state. The Zeeman field will orient the spin component and there will
be a residual $SU(2)$ valley symmetry. 
Introduction of anisotropies Eq.(\ref{aniso}) will have no effect on this eigenstate
since the wavefunction exactly vanishes when two electrons coincides because of 
Pauli principle and the model anisotropies involve only contact interactions. In 
the real world there will be also further anisotropies involving relative 
angular momentum one and more that will act upon the eigenstate. However these 
effects can be estimated as being $O(a/\ell)$ smaller than the contact anisotropies. 
However it is worth mentioning that theoretical estimates of anisotropies are 
much smaller than the values required to
explain the tilted-field transition between the F and CAF states at $\nu_G=0$. So 
this means that presumably also the anisotropies involving relative angular 
momentum one and higher are not well known and may be larger that naive estimates 
so they could possibly be relevant even in the range of filling factor $\nu<1$.

\subsection{The $\nu=2/3$ state}
\label{twothirds}
The situation is richer for $\nu=2/3$. 
Several competing states are known to be present at this filling factor. First of all 
there is
the one-component particle-hole symmetric of the $\nu=1/3$ state which is again a 
one-component state.
In the composite fermion description
this fully polarized state has negative effective flux and composite fermions 
occupy two effective $\Lambda$-levels~\cite{Jainbook}.
With two components one can also
construct a singlet state where now only one CF $\Lambda$ level is occupied 
by singlet pairs. For pure Coulomb interactions the singlet state 
has lower energy than the polarized state by approximately $0.009E_C$ in the 
thermodynamic limit. These two states compete directly on the torus geometry
while on the spherical geometry they have a different shift~: the polarized state
is realized for $N_\phi=(3/2)N_e$ and the singlet state for $N_\phi=(3/2)N_e-1$

While these two states are 
well-established in two-component FQHE systems, we note that there is 
evidence~\cite{Wu2015} for the 
formation of three-component and four-component states that are slightly 
lower in energy
by $\approx 0.002E_C$. These enigmatic states are not easily explained by 
composite-fermion theory and finite size limitations leas to a large uncertainly in
energy difference estimation. Such states are formed for $N_\phi=(3/2)N_e-2$
on the sphere and they also appear on the torus geometry.

All these pure Coulomb eigenstates can be embedded in the four-component case
giving rise to degenerate $SU(4)$ multiplets. Of course the $SU(4)$
singlet~\cite{Wu2015}
observed for $N_e=8$ is unaffected by anisotropies apart from a change in energy.
The polarized $2/3$ state does not feel the delta function interactions of 
Eq.(\ref{aniso}) but the singlet state has a nonvanishing probability of having 
two electrons at the same location provided they have different flavors~: 
$g_{\alpha\neq\beta}(0)\neq 0$. This probability is small and is known to be 
of the order of $10^{-3}$
from exact diagonalization or CF wavefunctions. As a consequence the splitting
induced by anisotropies is of order $g\times g_{\alpha\neq\beta}(0)$.
If we use the variational approach of section (\ref{irreps}) we obtain an energy 
functional
which has exactly the same expression as in the $\nu=2$ case apart from the 
overall scale.
This means that the phase diagram is the one given in Fig.(\ref{nu0phase}).
We have checked by exact diagonalization that this phase diagram is correct beyond perturbation theory. Notably all characteristics of the phase transitions are unchanged between $\nu=0$ and $\nu=2/3$. The ground state quantum numbers of the finite system
with $N=6$ and $N_\phi=8$ are consistent with the pattern of spin and valley ordering
for $\nu=2$.

\section{the fraction $\nu=4/3$}
We now turn to the richer situation with fractions appearing for filling
factors greater than one (and also less than two because of particle-hole symmetry).
At the fraction 4/3, by the remarks of section (\ref{irreps}) we  know that 
there are exact 
eigenstates in the $SU(4)$ symmetric limit obtained by adding a $\nu=1$ shell
to the $\nu=1/3$ polarized eigenstate of the Coulomb problem. The flavor content
of such a state is thus $(1,1/3)$. There is also an eigenstate obtained
by taking a $\nu=2/3$ singlet involving only two flavors and making a particle-hole 
transformation on both flavors so that its final
flavor content is $(2/3,2/3)$. 
With the known properties of the particle-hole transformation Eq.(\ref{pph}) 
in fact we already
know that the $(2/3,2/3)$ state has lower Coulomb energy than $(1,1/3)$ from the
energies of the parent states in the thermodynamic limit. This remark was made in 
ref.(\onlinecite{Inti}). Beyond these two exact eigenstates it is not guaranteed that 
there are not intruders implying more components. 

\subsection{The $(1,1/3)$ state}

In the torus geometry there is no shift so these two states directly compete
when we fix the flux and the number of particles, they differ only by the flavor
partitioning. In this geometry the state $(1,1/3)$ is an excited state
and it is thus computationally demanding to study it.

\begin{figure}[h]
\centering
 \includegraphics[width=0.8\columnwidth]{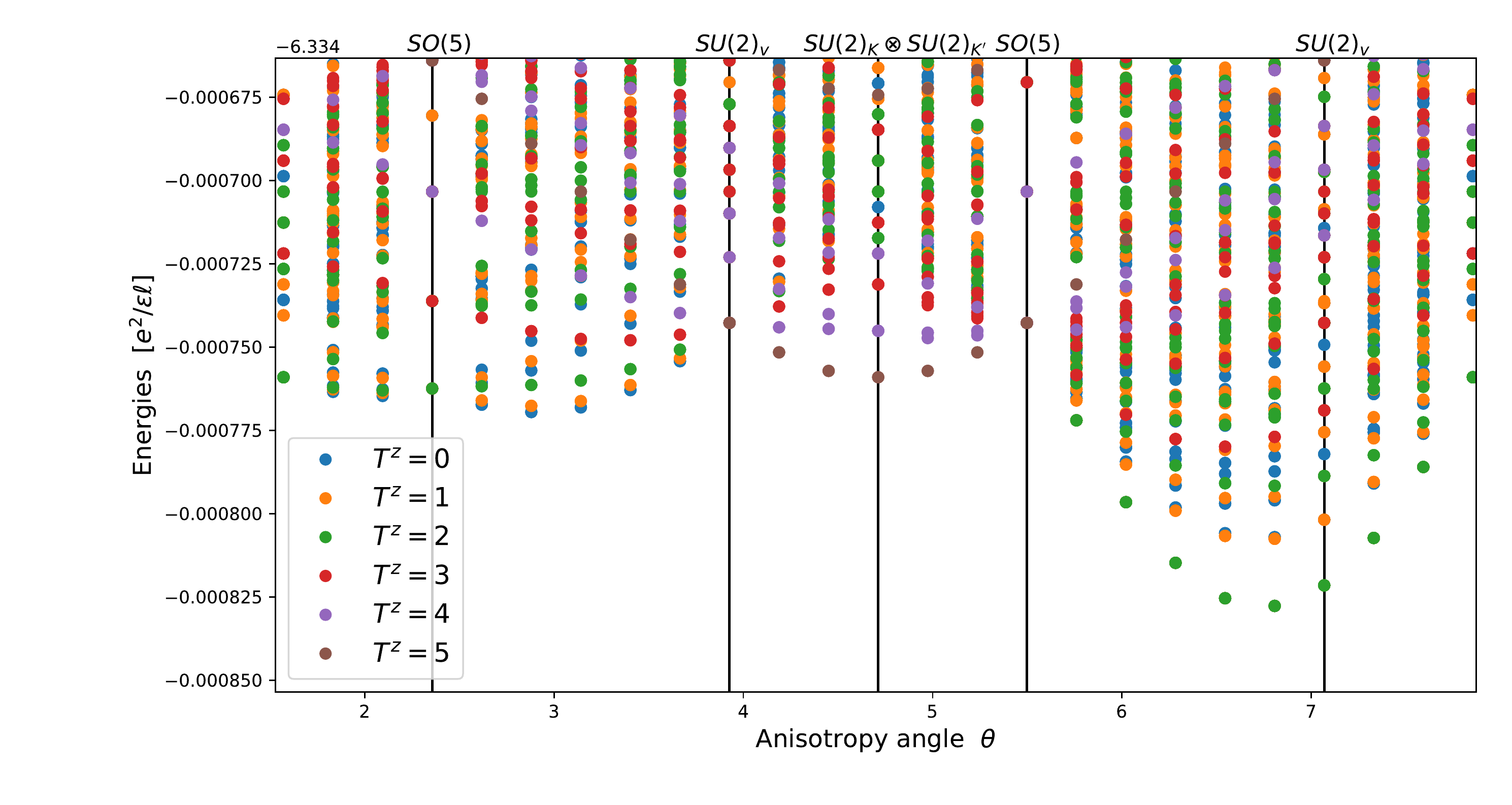}
 \caption{Energy levels as a function of anisotropy for $\nu=4/3$
 on the sphere geometry with $N_e=10$ and $N_\phi=6$. 
 The shift of the sphere geometry selects the 
 $(1,1/3)$ two occupied components of
 the highest-weight state.
 All these levels fan out from
 the parent unperturbed
 $SU(4)$ irrep which is $(7,3,0,0)$. In the $SU(4)$ symmetric limit this irrep is 
 not the ground state but it is the fourth excited state (fifth-lowest lying
 eigenstate). However it is the lowest-lying state with the expected
 quantum numbers for the $(1,1/3)$ state.
  The vertical lines mark the special symmetry points
 of the anisotropic interaction model we use. We identify four different phases.
 They can be called F.AF.KD and CDW as in the neutral case with the caveat
 that the quantum numbers do not match those predicted by simple trial wavefunctions.
 Notably the AF phase is valley unpolarized and spin singlet.
 Also the ground state of the F phase is fully spin polarized as expected
 but form $SU(2)$ valley multiplets with $S=5$ and $T=2$.
Unlike the $\nu=2$ case, phases AF and KD no longer
have the same quantum numbers and the phase transitions between them now involve 
a true level crossings at the $SO(5)$ point.}
 \label{sphere43_2C}
\end{figure}

In the sphere geometry
for the first candidate $(1,1/3)$ the total number of electrons is
partitioned
into two flavors $N_e=N_1+N_2$ and $N_1$ electrons fully fill
a spherical shell with flux $N_\phi$ while $N_2$ electrons form the usual 
$\nu=1/3$ state~:
\be
N_\phi = N_1 -1, \quad N_\phi= 3(N_2-1),
\ee
so that the flux-number of particles relationship is given by~:
\be
N_e=\frac{4}{3}N_\phi +2,
\ee
Hence we can use the sphere geometry to study separately the two states
$(1,1/3)$ and $(2/3,2/3)$.
We have performed exact diagonalizations of the system with $N_e=10$ and $N_\phi=6$.
While there is definitely a low-lying state with zero angular momentum $L=0$
and irrep $(7,3,0,0)$ as expected, it is not the ground state. Indeed the true ground state
spans the irrep $(4,4,1,1)$ with $L=0$ and the irrep $(7,3,0,0)$ is only the 
fourth excited state at this system size. At these rather small system sizes we consider that these states that lie below $(7,3,0,0)$
are likely quasiparticle states with flavor changing excitations.
The irrep $(7,3,0,0)$ is split by the anisotropies as shown in Fig.(\ref{sphere43_2C}).
We have used a small value ${\tilde g}=10^{-4}$ so that the states are not mixed with
the nearby irreps. 
The low-lying states centered onto the parent symmetric state $(7,3,0,0)$
are displayed in Fig.(\ref{sphere43_2C}). There are four distinct phases
separated by first-order transitions.  The range of existence of these four phases
is similar to the case for $\nu=2$. However the quantum numbers we find
are not always those predicted by the variational approach~:

\begin{itemize}
 \item 
For $-\pi/4<\theta<+\pi/2$ there is a phase with $S=5$ and $T^z=2$
as expected for a ferromagnetic state. Since the two valleys should be populated
by respectively seven and three electrons, one can have indeed the maximum
possible spin $S=5$ while $T^z$ is given by the difference in valley occupation.
However this is not the whole story since we observe that states with $T^z=0,1,2$
are exactly degenerate while the full Hamiltonian doe not have $SU(2)$ symmetry 
in the whole phase but only at the special point $\theta=\pi/4$.
This is not predicted by the wavefunction in Eq.(\ref{nu0wave}).

\item
For $+\pi/2<\theta<+3\pi/4$ we find a phase with $S=0$ and $T^z=0$,
which is natural to call antiferromagnetic. However it is definitely not
in agreement with the variational quantum numbers ($S=2$ and $T^z=2$).

\item
For $+3\pi/4<\theta<+5\pi/4$ the ground state has $S=2$ and $T^z=0$.
The value of $T^z$ points to XY valley order an these values are
those predicted variationally. So we call this phase  a K\'ekul\'e phase.
 
\item 
Finally  for $+5\pi/4<\theta<+7\pi/4$ we find $S=2$ and $T^z=5$.
The maximal value of $T^z$ means that all electrons reside in a single valley
i.e. a given sublattice
as in a charge-density wave state and the total spin is correctly predicted
by the variational wavefunction.
\end{itemize}
 
The most intriguing result is the appearance of exact valley multiplets
when the state is completely polarized. This can be explained by the following
line of reasoning~: when we have full spin polarization the electrons
occupy the two valleys and in the case of the $(1,1/3)$ state one of these valleys
is completely filled. If we perform a two-component particle-hole transformation
on the populated valleys we obtain a state $(0,2/3)$ which is fully polarized
since only one valley is occupied by holes. As a consequence there is no effect of
anisotropies since they involve only a contact interaction, requiring space
coincidence of electrons (as is the case of the $\nu=1/3$ state discussed 
in section (\ref{onethird})). This does not imply that such states have energies
independent of $g_z,g_\perp$ parameters because they appear in the one-body
terms in the particle-hole transformation.
So there is a subset of states that do not feel degeneracy-lifting anisotropies when
they are simply
given by contact interactions. Note that this argument is also valid for excited
states as long as they are also fully spin polarized
 
 The total spin and valley polarizations certainly lead us
 to name these phases as F,AF,KD,CDW as in the $\nu=2$ case.
 So the symmetry breaking pattern is the same as the neutrality case.
 However one of these four phases (AF) cannot be described by
 trial wavefunctions
in Eq.(\ref{product}) if we use for the spinors $\alpha$ and $\beta$ the spinors
that describe the $\nu=2$ phase diagram.
Another difference with respect to the neutrality case is that AF and KD phases do not 
have the same quantum numbers so the first-order transition
between them involves a ground state level-crossing unlike the $\nu=2$ 
case~\cite{Wu2014}.

\subsection{the $(2/3,2/3)$ state}

\begin{figure}[h]
\centering
 \includegraphics[width=0.8\columnwidth]{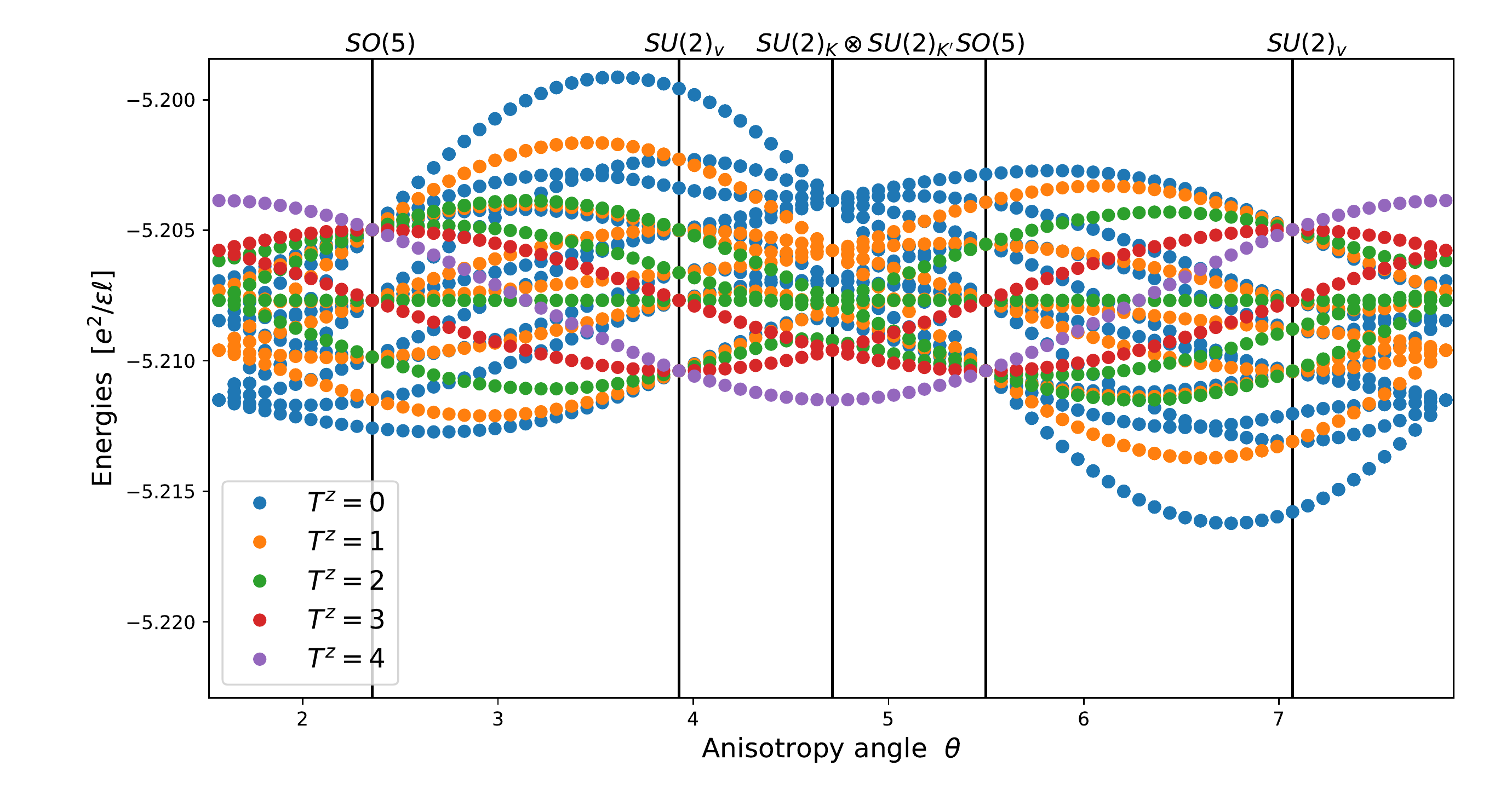}
 \caption{Energy levels as a function of anisotropy for $\nu=4/3$
 on the sphere geometry with $N_e=8$ and $N_\phi=5$. 
 The shift of the sphere geometry selects the 
 $SU(2)$ two-component singlet $(2/3,2/3)$ of the two occupied components of
 the highest-weight state.
 The parent unperturbed
 $SU(4)$ irrep is $(4,4,0,0)$ which is the ground state in the symmetric limit. 
  The vertical lines mark the special symmetry points
 of the anisotropic interaction model we use. We identify four different phases.
 The quantum numbers are exactly the same as in the neutrality case $\nu=2$.
 We thus observe that the transition between AF and KD phases does not involve
 a ground state level crossing but presumably happens through the collapse of
 a tower of states.}
 \label{Sphere_43}
\end{figure}

The well-known $SU(2)$ singlet state for $\nu=2/3$ is obtained  with a unit shift on the
sphere geometry
$N_\phi=(3/2)N_e-1$ with $N_e=N_\uparrow+N_\downarrow$. If we make 
the particle-hole transformation 
$N_{\uparrow,\downarrow}\rightarrow N_\phi+1 -N_{\uparrow,\downarrow}$
we obtain the relation $N_\phi=(3/4)N_e-1$. When embedded in the 4-component space
we expect to find an irrep with highest weight $(N_e/2,N_e/2,0,0)$ as the ground state.
The degeneracy is then lifted by the anisotropy~: in Fig.(\ref{Sphere_43}) we present results
of exact diagonalizations in the sphere geometry for $N_e=8$ electrons and $N_\phi=5$.
The important conclusion from this calculation is that the ground state quantum numbers
are now exactly the same as in the $\nu=2$ case.
So the phase diagram is the same as in the $\nu=2$ case~:  see Fig.(\ref{nu0phase})
with the same behavior at the phase transition points. Notably the AF/KD phase transition
has no ground state level crossing.

This is exactly what we find with the variational approach. Indeed with the
particle-hole symmetric of the singlet state we have now $g_{\alpha\beta}(0)$
of order unity and an energy functional Eq.(\ref{anisoenergypair}) equal to
that of $\nu=2$ except from an overall factor.

While the $(2/3,2/3)$ state is below the $(1,1/3)$ state at zero Zeeman energy
there may be a transition between these states that will be sensitive to the precise
phase which is realized. This is discussed in section (\ref{spintransition}).

\section{the fraction $\nu=5/3$}
Two possible candidates at this fraction are now $(1,2/3)$ which a two-component 
state and $(1,1/3,1/3)$ which is a genuine three-component state. 

\subsection{The two-component state $(1,2/3)$}
\begin{figure}[h]
\centering
 \includegraphics[width=0.8\columnwidth]{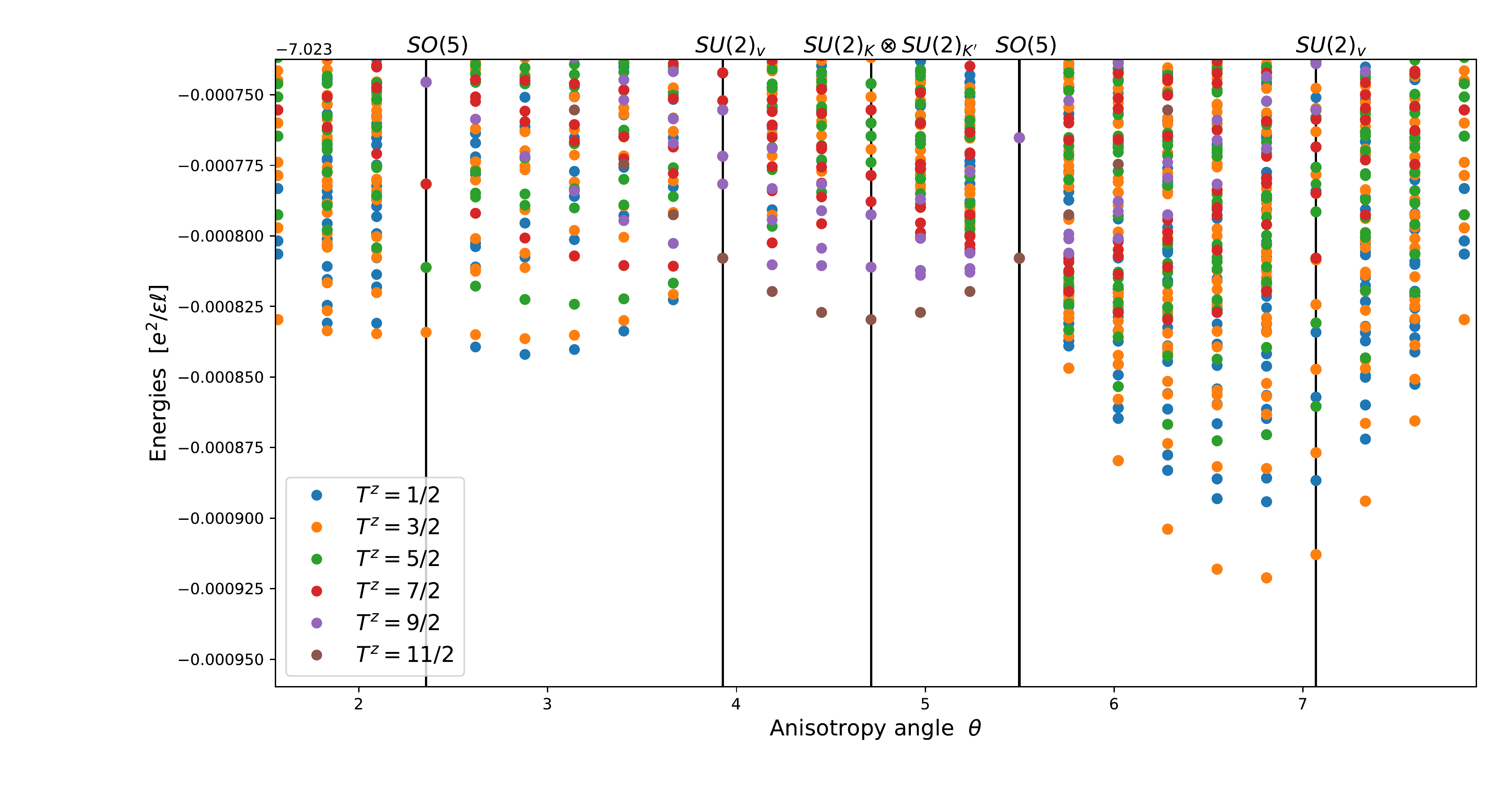}
 \caption{Energy levels versus anisotropy angle $\theta$ for filling factor $\nu=5/3$
 on the sphere geometry with $N_e=11$ and $N_\phi=6$ selecting the $(1,2/3)$ state.
 The parent $SU(4)$ irrep is $(7,4,0,0)$. This state is not the absolute ground state in the
 $SU(4)$ limit: it is the sixth excited state. One finds four phases
 consistent with the $\nu=2$ phase diagram but with distinct quantum numbers.
  There is a ferromagnetic phase for $-\pi/4<\theta<+\pi/2$ with $S=11/2$ and $T^z=1/2,3/2$ so there is an emergent $SU(2)$ valley symmetry,
 an antiferromagnetic phase for $+\pi/2<\theta<+3\pi/4$ with $S=T^z=3/2$,
 a K\'ekul\'e phase for $+3\pi/4<\theta<+5\pi/4$ with $S=3/2$ and $T^z=1/2$
 and a charge-density wave phase for $+5\pi/4<\theta<+7\pi/4$ with $S=3/2$ and $T^z=11/2$.
 All transitions are first-order with level crossings.}
 \label{Fivethird2C}
\end{figure}

The first state 
is obtained by adding the polarized
i.e. one-flavor $\nu=2/3$ state to a filled level. This polarized state with $\nu=2/3$
is the one-component particle-hole transform of the polarized Coulomb eigenstate
at $\nu=1/3$. The flux-number of particles is thus given by~:
\beb
N_\phi = N_1 -1, \quad & &N_\phi= 3/2 \times N_2,\quad N_e=N_1+N_2,\\
N_\phi&=&\frac{3}{5}(N_e-1).
\eeb
We have performed sphere exact diagonalizations for $N_e=11$ and $N_\phi=6$.
In the $SU(4)$ limit there is a lowest-lying state with $(7,4,0,0)$ irrep and $L=0$
as expected for the state obtained from the gluing procedure of Eq.(\ref{product})
but it is not the ground state. This is the same phenomenon that we observe at 
$\nu=4/3$.
We posit that these extra states are flavor-changing quasiparticle excitations
and focus only onto the fate of the $(7,4,0,0)$ multiplet. By using a small value of 
the anisotropy ${\tilde g}=10^{-4}$ the irrep is split in many levels but they do not
mix with
other multiplets. The result of this calculation is displayed in 
Fig.(\ref{Fivethird2C}).

As in the case of the $(1,1/3)$ state the quantum numbers are exactly those
expected from using the spinors $\alpha$ and $\beta$ describing the various orderings
of $\nu=2$ F,AF,KD,CDW and using them with Coulomb eigenstates in Eq.(\ref{product}).
Also since AF and KD
do not have the same quantum numbers there is a level-crossing phase transition at 
the $SO(5)$ point between AF and KD phases.

In the fully polarized sector we observe also the appearance of degeneracies
due to the $SU(2)$ valley symmetry not only at the special point $\theta=\pi/4$
but in the whole F phase. The manifold of $S=11/2$ states involves indeed 
$T^=1/2$ and $T^z=3/2$
while the variational prediction is that we should observe only $T^z=3/2$.
This is due to the same phenomenon we found for the
ferromagnetic phase in the case of the $(1,1/3)$ state. The two-component 
particle-hole symmetric
state of the fully polarized states has valley content $(0,1/3)$ so the holes being
polarized do not feel the point-contact anisotropies.

\subsection{the three-component state}
For the three-component state one now replaces the $\nu=2/3$ polarized state
by the two-component singlet state at the same filling factor leading to
a flavor content $(1,1/3,1/3)$. We call $N_1$ the number of electrons in the fully 
filled LL
and $N_2,N_3$ the electron numbers in the two flavors forming the singlet state.
We obtain thus a shift on the sphere which is different from that of the previous state~:
\beb
N_\phi = N_1 -1, \quad & &N_\phi= 3/2 \times (N_2+N_3)-1,\quad N_e=N_1+N_2+N_3,\\
N_\phi&=&\frac{3}{5}N_e-1.
\eeb
We have studied in detail the case $N_e=10$, $N_\phi=5$.
In the $SU(4)$ limit we are already certain that
this state is lower in energy than $(1,2/3)$ since the $\nu=2/3$ singlet is lower
in energy than the polarized state at the same filling factor. We are also certain that
there is such an eigenstate of the $SU(4)$ symmetric Coulomb problem. 

The only
remaining question is if there are some states lower in energy. Indeed it may very well
be that by spreading the electrons into more flavors one can reduce the energy cost of
Coulomb repulsion. If we consider the possible existence of an $SU(3)$ singlet state
at filling $2/3$ with shift 2 on the sphere~\cite{Wu2015} then it implies the existence
of a state with flavor content $(1,2/9,2/9,2/9)$ and a flux given by 
$N_\phi=(3/5)N_e -(7/3) $. We observe on the sphere geometry that the ground state
for $N_e=14$ and $N_\phi=7$ is spanned by the irrep $(8,2,2,2)$.
Since we already know~\cite{Wu2015} that there is a $SU(3)$ singlet $(2,2,2)$ for 
$N_e=6$ and $N_\phi=8$
this is in fact only a consistency check. Due to the severe size limitations of exact
diagonalizations we cannot shed further light on this issue and limit ourselves
to the states $(1,2/3)$ and $(1,1/3,1/3)$. If there are states like $(1,2/9,2/9,2/9)$
they are relevant only at small Zeeman energy.

\begin{figure}[h]
\centering
 \includegraphics[width=0.8\columnwidth]{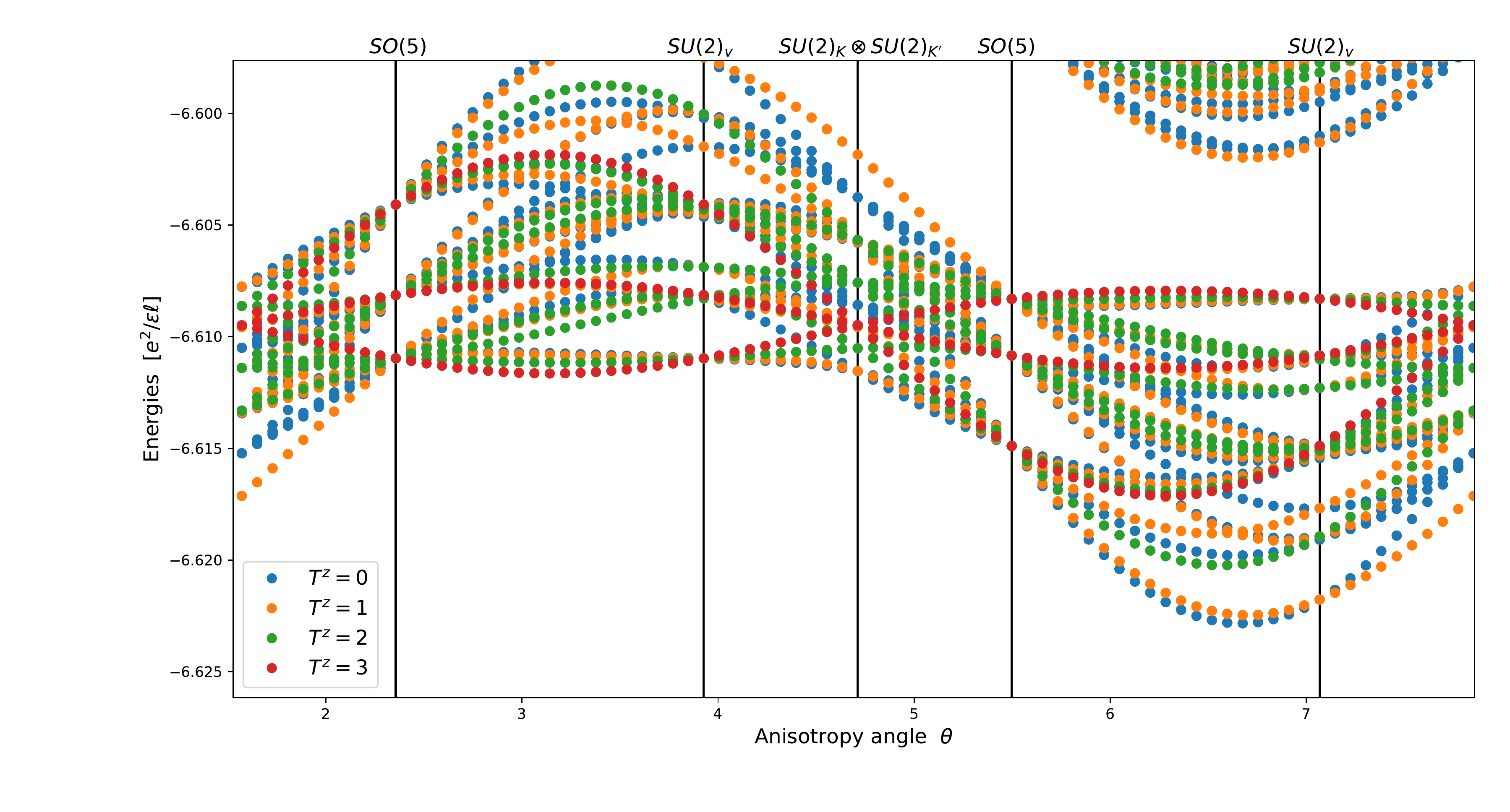}
 \caption{Energy levels versus anisotropy angle $\theta$ for filling factor $\nu=5/3$
 on the sphere geometry with $N_e=10$ and $N_\phi=5$. This special shift
 favors the singlet state for the two partially occupied Landau levels
 so that the flavor content is $(1,1/3,1/3)$.
 The parent $SU(4)$ irrep is $(6,2,2,0)$ and it is the ground state. 
 The magnitude of the anisotropy
 is ${\tilde g}=10^{-2}$. We observe five phases whose quantum numbers are displayed
 in Fig.(\ref{Fivethirdphase}). The five quantum phase transitions are first-order
 with true level crossings.}
 \label{Sphere_10_5}
\end{figure}

\begin{figure}[h]
\centering
 \includegraphics[width=0.3\columnwidth]{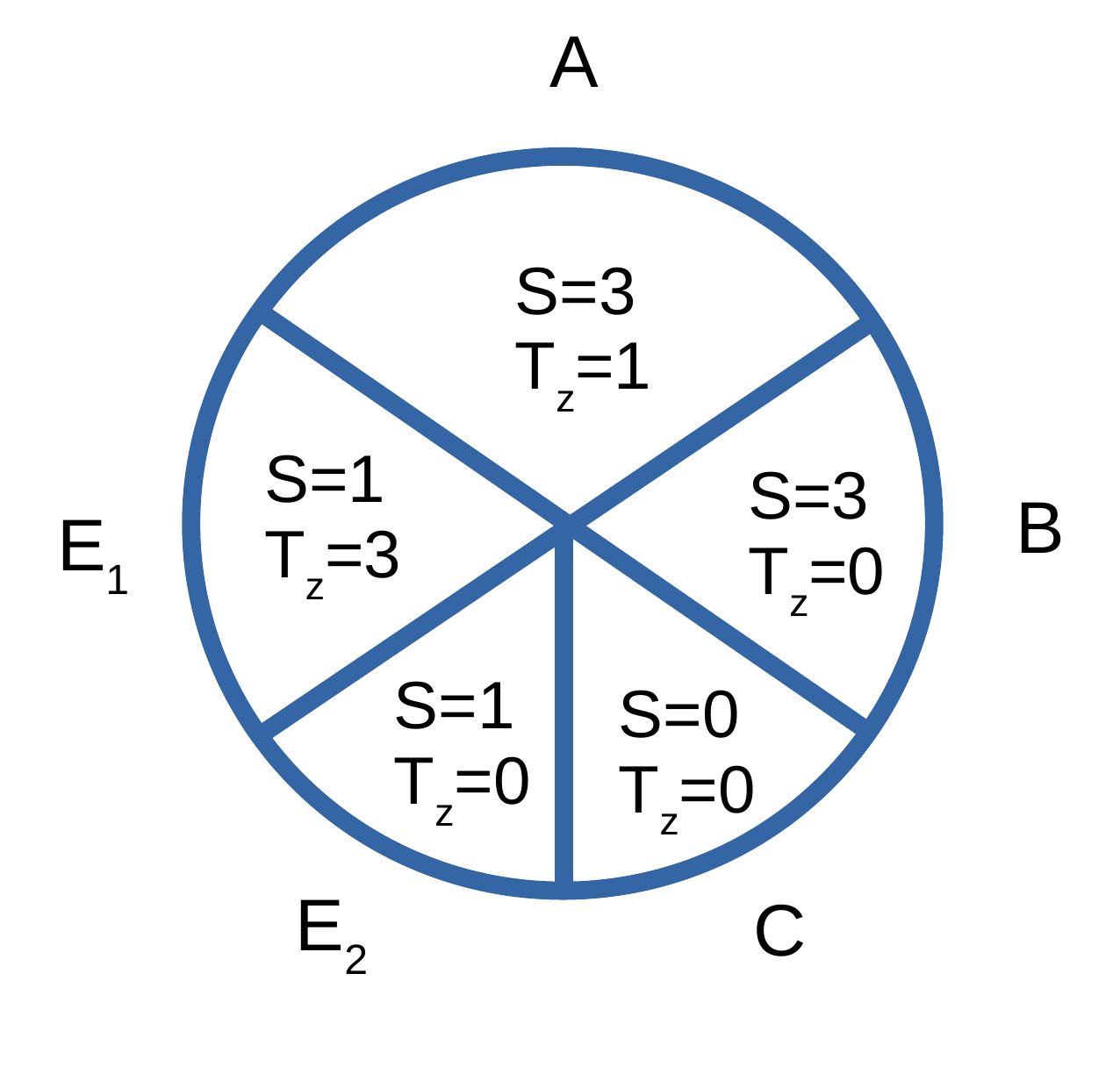}
 \caption{Phase diagram for the $\nu=5/3$
  with $N_e=10$ and $N_\phi=5$.
 The parent $SU(4)$ irrep is $(6,2,2,0)$. This is valid on the sphere geometry. 
 On the torus geometry the quantum numbers are slightly different
 since the number of states per Landau level differ by one unit.
 While $A$,$B$ can be captured plausibly by the variational method, it is not case of the singlet phase $C$. The $E_{1,2}$ phase are found to be degenerate variationally while
 our results show that they differ by type of valley ordering.}
 \label{Fivethirdphase}
\end{figure}

Our exact diagonalization results on the sphere geometry are presented in 
Fig.(\ref{Sphere_10_5}).
We find now a phase diagram \textit{different} from the neutrality case.
There are five phase transitions and all of them involve level crossings in
the finite systems.
The five phases we observe have quantum numbers displayed in 
Fig.(\ref{Fivethirdphase}).
Some of them may be captured by the variational approach but not all.

Since we are dealing with a three-component state it is no longer possible
to have full spin or valley polarization. The maximal value of the spin is observed
in the range $-\pi/4<\theta < +3\pi/4$. In this regime we observe two phases $A$ 
and $B$ 
that differ by the valley $T^z$ value. 

\begin{itemize}

\item
The $A$ phase has no valley XY order and can be described by
$\{\alpha,\beta,\gamma\}=\{|K \uparrow\rangle,|K^\prime \uparrow\rangle,
|K^\prime \downarrow\rangle\}$ 
with $T^z=1$ and $S=3$.

 \item 
The $B$ phase or $-\pi/4<\theta < +\pi/4$
has $T^z=0$ and is plausibly described by a state with 
$\{\alpha,\beta,\gamma\}=\{|\mathbf{t}_\perp \uparrow\rangle,
|-\mathbf{t}_\perp \uparrow\rangle,
|-\mathbf{t}_\perp \downarrow\rangle\}$. 
Due to the three-component nature of the state
it is not possible to obtain $T^z=0$ by using states with definite projections 
onto $|K\rangle$,
$|K^\prime\rangle$ but one has to use XY valley ordered states.

\end{itemize}

The transition between $A$ and $B$ phases is thus associated to the
change of the valley order from Ising-type in $A$ to XY-type in $B$.

The region $3\pi/4<\theta < 3\pi/2$ has now only partial spin polarization
and is divided into two phases that differ by a change in the value of the 
valley polarization. 

\begin{itemize}

 \item 
We find a phase $E_1$ for $3\pi/4<\theta < 5\pi/4$
with the maximal value of $T^z$.
$\{\alpha,\beta,\gamma\}=\{|K \uparrow\rangle,|K \downarrow\rangle , |K^\prime \downarrow\rangle\}$

\item
In the lower quadrant $5\pi/4<\theta < 3\pi/2$ we have a phase $E_2$
with $T^z=0$ indicative of XY valley ordering whose candidate ordering pattern
is given by
$\{\alpha,\beta,\gamma\}=\{|\mathbf{t}_\perp \uparrow\rangle,|-\mathbf{t}_\perp \downarrow\rangle,
|\mathbf{t}_\perp \downarrow\rangle\}$.
The transition between $E_1$ and $E_2$ corresponds in changing the valley order
from the Ising-like $z$-axis to the valley XY plane.
A variational treatment does not distinguish between Ising or XY character
in this range of anisotropies. Indeed all states are degenerate with trial energy
in Eq.(\ref{anisoenergypair}).

\item
Finally there is a fifth phase that we call $C$ for $3\pi/2<\theta < 7\pi/4$
which is a spin singlet $S=0$ as well as valley unpolarized $T^z=0$.
It is not possible to capture such a phase with the class of 
variational states discussed above. If we look at the full set of degenerate states
in the $SU(4)$ limit, one notes that the irrep $(6,2,2,0)$ that we study
contains notably symmetric states like $(3,2,2,3)$ from which one can
construct states with zero spin and zero isospin values. It is an open question
to obtain explicitly a wavefunction with the correct quantum numbers for the $C$ phase.
\end{itemize}

\begin{figure}[h]
\centering
 \includegraphics[width=0.5\columnwidth]{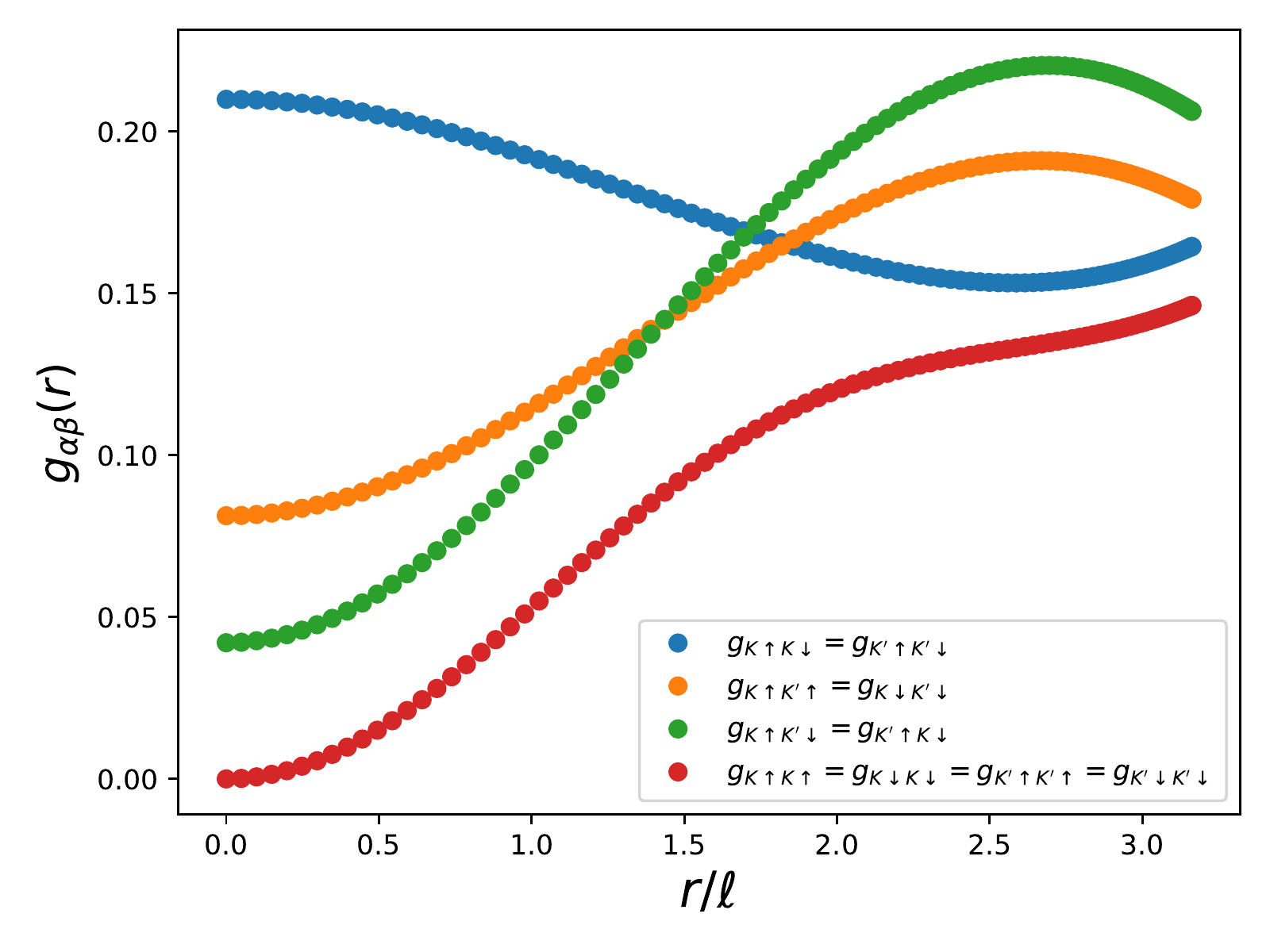}
 \caption{The various pair correlation functions $g_{\alpha\beta}(r)$
 calculated in the middle of the singlet $C$ phase for $\theta=3\pi/2+\pi/8$.
 The sphere geometry is used with $N_e=10,N_\phi=5$. The chord distance $r$ varies from zero 
 up to $\sqrt{10}\ell$.
 Since we have $S^z=0$ and $T^z=0$ in this phase there are only four distinct
 correlations.}
 \label{paircorrelation}
\end{figure}

To shed some more light onto the nature of the $C$ phase we have computed the pair
correlation function $g_{\alpha\beta}(r)$ of the exact ground state in the sphere
geometry. With a ground  state having zero spin and $T^z=0$ there are only four
independent combinations of spin-valley that are plotted in Fig.(\ref{paircorrelation})
At short distance the leading correlation is 
$g_{K\uparrow K\downarrow}(0)=g_{K^\prime\uparrow K^\prime\downarrow}(0)$.
This function has a maximum at the origin while all other cases have a deep
minimum as expected from Coulomb repulsion.
This may point to formation of spin S=0 singlet pairs in each valley.
On the contrary the antiferromagnetic-like repulsion between
$K\uparrow$ and $K^\prime\downarrow$ is maximal at a finite distance
$\approx 2.5\ell$.

\section{spin transitions}
\label{spintransition}

In the case of filling factor $\nu=2/3$ it is well known~\cite{Ashoori} that 
one can induce a
spin transition between the singlet state and the fully polarized state. 
Indeed
while the polarized state has a higher Coulomb energy, increasing the Zeeman energy will lower
it eventually below the singlet state. The crossing happens when the Zeeman energy
equals the energy difference between the two states~:
\be
\Delta E =\epsilon_Z B_{crit}.
\ee
This prototypical transition is simple because of the well-defined magnetization of 
the two competing states. Here in monolayer graphene  the situation is richer
since the competing states that we have studied above can have various magnetizations
according to the value of the anisotropy parameters. The Coulomb energy
scales as $\propto \sqrt{B}$ while anisotropy energies and the Zeeman energy
scale linearly with $B$. The energy per particle of a state $i$ has thus three
contributions~:
\be
\epsilon_i = E_{Ci} + a_i B + \epsilon_Z m,
\ee
where we have defined the magnetization per particle $m$
and the Coulomb energy $E_{Ci}$
A spin transition between two states $0$ and $1$ arises when the Coulomb energy difference is equal
to the contribution from anisotropies and Zeeman energy~:
\be
\Delta\epsilon_{01}(B_{crit}) = (a_0-a_1+z)B_{crit}
\label{criticalfield}
\ee
The Zeeman factor $z$ depends upon the magnetization of the two competing states
and in the case of monolayer graphene it takes different values in different parts of
anisotropy phase diagram.

We now give a simplified description of spin transitions for the fractions we have
studied. We limit ourselves to situations with only a perpendicular field
and also we ignore the possibility of spin canting. Indeed the effect of the canting
is restricted to low enough fields. At large enough values of the field one obtains
a fully polarized state or a collinear antiferromagnet depending on the filling 
fraction. Certainly spin canting may drive interesting transitions but detailed
predictions are hampered by our lack of knowledge about the values of anisotropies.

\subsection{$\nu=2/3$}
At filling factor $2/3$ the polarized state is insensitive to anisotropies
while the singlet state can be in any of four phases F,AF,CDW,KD
as shown in section (\ref{twothirds}).
In KD or CDW the state is a spin singlet so it is insensitive to the Zeeman coupling.
We thus expect a spin transition towards the fully polarized state
at some field value Eq.(\ref{criticalfield}). The AF state turns into a canted 
antiferromagnetic state which becomes fully polarized beyond some field value.
Once in this fully polarized state, as is the case of the F phase, the state lowers 
its energy
at the same rate as the polarized $\nu=2/3$ state so that there will be no crossing
hence there will be no spin transition in the AF or F phases.

\subsection{$\nu=4/3$}
The phase diagrams for the states $(1,1/3)$ and $(2/3,2/3)$ are similar~:
there is the same number of phases with their domain of stability having 
the same range of anisotropy angle $\theta$.
but the difference is now in their total magnetization.
In the F phase the two states have the same total spin value so they never
cross under Zeeman coupling. In the KD and CDW phases the situation is different.
In the state $(2/3,2/3)$ there is zero net magnetization in both cases KD and CDW
while in the state
$(1,1/3)$ the KD phase has now a net magnetization equal to 1/3 of the saturation value
and the CDW phase has an even smaller magnetization (but nonzero).
So we deduce that there can be a spin transition in both phases KD and CDW.

\subsection{$\nu=5/3$}
The situation is more complex since now the phase diagrams
of the two competing states $(1,2/3)$ and $(1,1/3,1/3)$ do not overlap exactly
as a function of anisotropy.
We describe the situation by using as a reference the phases of the $(1,2/3)$ state.
In the F phase the state $(1,2/3)$ is fully polarized with magnetization
$M=M_{sat}$ while $A$ and $B$ phases of $(1,1/3,1/3)$ have only $M=M_{sat}/\nu$
so we expect a spin transition.
In KD and CDW phases the higher-lying state $(1,2/3)$ has $M=M_{sat}/3$
while the competing phases in $(1,1/3,1/3)$ are the $C$ and $E_2$ phases 
(see Fig.(\ref{Fivethirdphase})).
In the singlet $C$ phase we have $M=0$ and in the $E_1$ and $E_2$ phases $M=M_{sat}/5$
so there will be a spin transition. Its location $B_{crit}$
will be phase dependent because there are (yet unknown) contributions from
anisotropies in the value of the critical field in Eq.(\ref{criticalfield}).

The general picture is that states with maximal spread-out of electrons in various 
spin-valley components will be favored only at small Zeeman energies.
Notably graphene experiments with large Zeeman splittings and large sublattice
effects like measurements in ref.(\onlinecite{Zeng-MLG9}) will involve only states like $(1,1/3)$
and $(1,2/3)$ as well as their generalizations to other fractions.
Genuine multicomponent states will require minimal effect of the one-body fields.

\section{conclusion}

We have studied the impact of anisotropies relevant to the description
of monolayer graphene in the regime of the fractional quantum Hall effect.
At neutrality the phase diagram involves four phases F, AF, KD and CDW.
Simple generalizations of this diagram apply for $\nu=(1/3,1/3)$,
$\nu=(2/3,2/3)$
and $\nu=(1,2/3)$ states. This is found from exact diagonalizations on the 
sphere geometry and is also confirmed by a variational approach involving parent
Coulomb eigenstates. The two spin-valley vectors $\alpha,\beta$ that characterize
the spin-valley order in the variational approach are identical to the neutral case.
Since the occupations of the two filled states are in general not equal
it means that the quantum numbers of the ground state are now different from the
neutral
case. As a consequence all first-order transitions involve 
level crossings. Indeed there is no evidence for exotic phase transitions\cite{Subir}.

In the case of $\nu=(1,1/3)$ there are also four phases whose range of stability
is the same as the neutral case but the quantum numbers are not all predicted
by the variational method. While CDW and KD-like phases have spin and valley
quantum numbers correctly predicted, we find that the antiferromagnetic phase
is a spin singlet with no net valley polarization. Interestingly we
find that the fully polarized states partly escape effects of anisotropy and still
form $SU(2)$ valley multiplets even though this is not a symmetry of the 
Hamiltonian (their energies still depend upon $g_z$ and $g_\perp$).
This emergent symmetry also appears in the polarized eigenstates of the $(1,2/3)$
state.

The case $\nu=(1,1/3,1/3)$ is different. We observe five phases. Two of them
can be described variationally. There are two distinct phases in our diagonalizations
that differ by Ising valley order versus XY valley order while they are degenerate
variationally. There is also a phase which a spin singlet with presumably XY valley
order that happens for negative Ising-like anisotropy. It is an open question
how to write a wavefunction to describe this phase. The pair correlation function
$g_{\alpha\beta}(0)$
shows an enhanced probability for electrons for opposite spins but in the same
valley to be at the same location. This interesting situation requires
however low Zeeman energy to be realized experimentally.

In present experiments it is likely that one observes the two-component states
$(1,1/3)$ and $(1,2/3)$. If they are fully polarized (the F phase) then we predict
that they should escape degeneracy-lifting
anisotropies of the form given in Eq.(\ref{aniso})
and thus feature an emergent valley $SU(2)$ symmetry. Invalidating this valley
symmetry would invalidate the simplified model of Eq.(\ref{aniso})
which is a crucial piece of our present understanding of IQHE and FQHE in graphene
systems.

Recent experiments using scanning tunneling microscopy~\cite{Yazdani,Sacepe}
have given evidence for a more complex picture at neutrality $\nu=0$ than previously
thought. Notably there is evidence for phases beyond the four states of 
Fig.(\ref{nu0phase}).
There are at least two possible explanations. It may very well be that
the anisotropies are not small in comparison to coulomb and that the simple
model Eq.(\ref{aniso}) is not adequate. It may also mean that the Landau level 
mixing
is strong enough to change the phase structure. Such a possibility would invalidate
standard theoretical treatments that focus on a fixed Landau level from the start.
Also some of the experiments~\cite{Yazdani} favor a K\'ekul\'e state which is
at odds with the explanation of the metal-insulator transition~\cite{Young-MLG2}.
This may mean that the anisotropy parameters are not in the range of 
Eq.(\ref{anisobounds}) in which case the explanation for the titled-field transition
becomes more elusive. It may be that the edge of the sample
lies in a different phase from the bulk as observed in Hartree-Fock
studies~\cite{AK}.

\begin{acknowledgments}
We acknowledge discussions with A. Assouline and P. Roulleau. 
We thank C. Repellin for useful correspondence.
We thank DRF and GENCI-CCRT for 
computer time allocation on the Cobalt and Topaze clusters.
\end{acknowledgments}


\end{document}